\documentclass[a4paper]{article}

\usepackage{ISCSLP2022}
\usepackage{makecell}
\usepackage[skip=3pt]{caption} 
\usepackage{multirow}
\usepackage{cite}
\usepackage{amsmath}
\usepackage{url}

\usepackage{breakurl}

\title{Label-free Knowledge Distillation with Contrastive Loss for Light-weight Speaker Recognition}
\name{Zhiyuan Peng$^{1,2}$, Xuanji He$^2$, Ke Ding$^2$, Tan Lee$^1$, Guanglu Wan$^2$}
\address{
  $^1$Department of Electronic Engineering, The Chinese University of Hong Kong\\
  $^2$Meituan}
\email{jerrypeng1937@gmail.com, \{hexuanji,dingke02,wanguanglu\}@meituan.com, tanlee@ee.cuhk.edu.hk}

\begin{document}

\maketitle
\begin{abstract}
  Very deep models for speaker recognition (SR) have demonstrated remarkable performance improvement in recent research. However, it is impractical to deploy these models for on-device applications with constrained computational resources. On the other hand, light-weight models are highly desired in practice despite their sub-optimal performance. This research aims to improve light-weight SR models through large-scale label-free knowledge distillation (KD). Existing KD approaches for SR typically require speaker labels to learn task-specific knowledge, due to the inefficiency of conventional loss for distillation. To address the inefficiency problem and achieve label-free KD, we propose to employ the contrastive loss from self-supervised learning for distillation. Extensive experiments are conducted on a collection of public speech datasets from diverse sources. Results on light-weight SR models show that the proposed approach of label-free KD with contrastive loss consistently outperforms both conventional distillation methods and self-supervised learning methods by a significant margin.
\end{abstract}

\section{Introduction}\label{sec:intro}

Speaker embeddings learnt by deep neural networks have shown very impressive performance for speaker recognition (SR)\cite{snyder2018x,zeinali2019but,desplanques2020ecapa}. It has been extensively verified that neural models with deeper residual layers (such as ResNet\cite{he2016deep,zeinali2019but}) and more complex neural blocks (such as ECAPA-TDNN\cite{desplanques2020ecapa}) can significantly outperform the classical x-vector\cite{snyder2018x} extractor.
However, these models are computationally demanding. It is difficult, if not impossible, to deploy them for resource-constrained applications that have limited computational budgets and require low real time factor.

The task of developing light-weight models for on-device application has drawn a lot of attention. A common methodology is to explore more efficient neural network architectures, e.g., MobileNet\cite{howard2019searching}, ShuffleNet\cite{ma2018shufflenet}, EfficientNet\cite{tan2019efficientnet}, etc.
Another direction is to compress neural network models through parameter pruning\cite{han2015deep,han2015learning}, quantization \cite{gong2014compressing,wu2016quantized,zhu2021binary}, and low-rank decomposition\cite{denton2014exploiting,lebedev2014speeding,georges2020compact}. The light-weight models developed with these approaches have small model size and low latency for inference. They are suitable for on-device deployment, but often at the expense of performance degradation.

To improve the performance of light-weight models, the knowledge distillation (KD) approach has been investigated\cite{wang2019knowledge,georges2020compact,lin2021towards}. For speaker recognition, many well-trained speaker extractors have been developed \cite{pretrained2017xvector,pretrained2020chung,pretrained2018xvector}.
KD is to distill useful knowledge from these excellent models (teacher) to improve light-weight extractor models (student). A teacher model maps speech samples into speaker-discriminative embeddings. The embeddings are then leveraged as a kind of supervision to train student models. Through this process, student models can be improved by benefiting from both a large amount of data and superior teacher models. In the context of pre-training for light-weight models, KD could be both efficient and effective. Related research in this direction is quite limited. To our knowledge, only Georges\cite{georges2020compact} and Wang\cite{wang2019knowledge} considered applying KD to light-weight speaker embedding extraction.

With the motivations stated above, the present study targets on improving light-weight speaker extractors with label-free KD on large amounts of unlabelled data. 
The major innovation of the proposed label-free KD is the design of training loss.
Conventional KD defines the training loss as an interpolation between the distillation loss and an auxiliary classification loss.
The classification loss is generally necessary for learning task-specific knowledge, but at the expense of golden labels.
This is because of the inadequacy in teacher-student learning by the conventional distillation loss\cite{wang2019knowledge}, as will be evidenced in our experiments. To handle the insufficiency issue while discarding the classification loss, we leverage the contrastive loss \cite{he2020momentum} from self-supervised learning (SSL).
We experimentally find that KD along with the contrastive loss can significantly improve the performance of light-weight models. It can be even better than 
the state-of-the-art angular-based classification loss\cite{deng2019arcface,liu2019large} that requires speaker labels. As a result, label-free KD can be achieved.

Label-free KD can also be used for unsupervised domain adaptation. Experiments show that when adapting to data with severe domain mismatch, KD with contrastive loss can outperform conventional distillation methods by a significant margin. The performance is also comparable to (sometimes better than) supervised training with out-of-domain data.

Finally, we scale up the amount of unlabelled data to carry out large-scale label-free KD.
Several large-scale datasets collected from different speech tasks are combined together, reaching a total duration of 24,182 hours after pre-processing. Large-scale KD on the combined data can further boost the performance of light-weight speaker extractors. It is worth noting that since the extractors are light-weight, large-scale KD is actually computationally affordable to the academic community.

\section{Label-free Knowledge Distillation}
% describes the structures of teacher-student modeling without labels with a picture
\begin{figure}[htbp]
    \centering
    \includegraphics[width=0.3\textwidth]{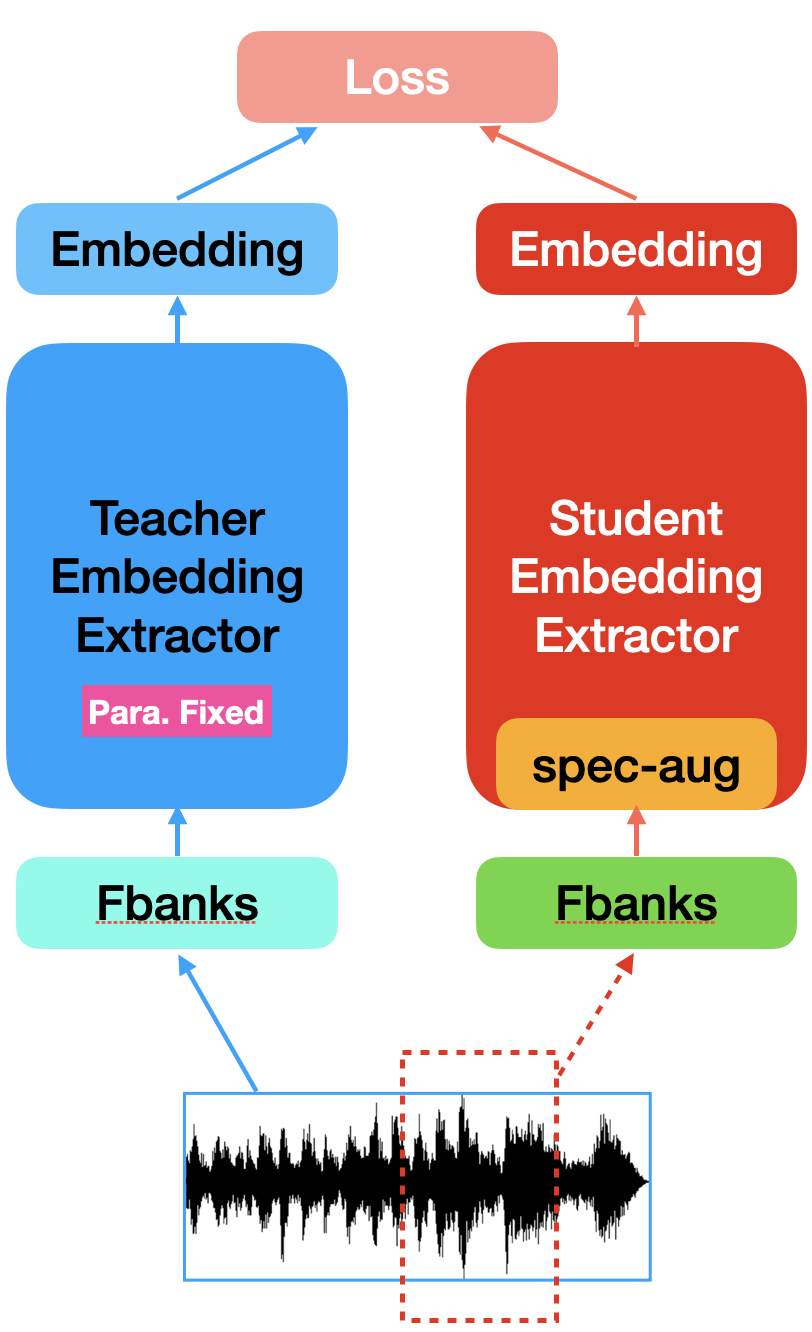}
    \caption{teacher-student training for label-free knowledge distillation}
    \label{fig:structure}
\end{figure}

Similar to the previous work\cite{wang2019knowledge}, the teacher-student structure is adopted for achieving KD. As shown in Figure \ref{fig:structure}, given an input speech waveform, the teacher embedding extractor uses filter-bank features (fbanks) of the whole input to generate a per-utterance speaker embedding. A short clip (2s-3s) of the waveform is cropped at a random time point in the utterance to generate fbanks as the input to the student embedding extractor. 
In addition, spectrum augmentation\cite{park2019specaugment} with time-frequency masking and warping of the input fbanks is applied to the student embedding extractor.
% The two extracted embeddings are then used to compute a distillation loss that measures the distance of embeddings.
The teacher model is assumed to be well-trained and its parameters are fixed in KD.
The parameters of the student model are randomly initialized and updated by minimizing a distillation loss.
The distillation loss is defined as the distance between pairs of embedding generated from the teacher and student models.
Through this distillation process, it is expected that the student can learn adequate knowledge from the teacher to generate discriminative speaker embeddings.

A critical difference between our work and the previous\cite{wang2019knowledge} is the training loss. The existing distillation method typically employs an auxiliary classification loss. The overall training loss is defined as a weighted combination of the distillation loss and the classification loss. This is because using the distillation loss only may not be sufficient for effective learning of task-specific knowledge. The interpolation could to a certain extent improve the performance of student learning. However, as labelled data are needed for the classification loss, the amount of data available for distillation would be limited. The present study aims to get rid of the classification loss and improve the distillation loss so as to unleash the power of large-scale distillation on unlabeled data.

\subsection{Distillation Loss\label{sec:distillation_loss}}
% Most speaker verification systems require a superior speaker extractor that generates discriminative speaker embeddings. Therefore 
We consider distillation at the embedding level, as shown in Figure \ref{fig:structure}. Two simple loss functions were proposed in the previous work\cite{wang2019knowledge}. They are the mean square error (\textit{MSE}) and the cosine distance (\textit{COS}) loss,
\begin{equation}
    \mathcal{L}_{\text{MSE}} = \sum_{i=1}^{N} ||v_t^i - v_s^i ||^2
\end{equation}
\begin{align}
    &\mathcal{L}_{\text{COS}} = -\sum_{i=1}^{N} \cos{(v_t^i, v_s^i)} \\
    &\cos(v_t^i, v_s^i) = \frac{<v_t^i, v_s^i>}{||v_t^i||_2||v_s^i||_2}
\end{align}
where $v_t^i$ and $v_s^i$ denote the embeddings extracted from the teacher and the student for the $i$-th speech utterance. $N$ refers to the batch size. 

\subsubsection{Contrastive loss}
% These loss functions are still incapable of distilling sufficient knowledge from the teacher such that an additional classification loss is also required in the training.
To improve the distillation performance for SR, we adopt the contrastive loss\cite{he2020momentum},
\begin{equation}
    \mathcal{L}_{\text{contra}} = -\sum_{i=1}^{N} \ln \frac{\exp{(\cos(v_t^i, v_s^i) /\tau) }}{\sum_{j=1}^N \exp{ ( \cos(v_t^i, v_s^j)/\tau)}}
\end{equation}
where $\tau$ is the temperature that scales up the cosine similarities to logits before softmax normalization. Empirically, we find that $\tau = 0.1$ works pretty well.

For the $i$-th teacher-student embedding pair, the \textit{COS} loss measures the similarity $\cos(v_t^i, v_s^i)$ between $v_t^i$ and $v_s^i$. The contrastive loss, on the other hand, incorporates a denominator term $\sum_{j=1}^N \exp{ (\cos(v_t^i, v_s^j)/\tau)}$ to normalize the similarity. The denominator enumerates the pairwise similarities $\cos(v_t^i, v_s^j)$ between all the student embeddings $\{v_s^j\}_{j=1}^N$ and the $i$-th teacher embedding $v_t^i$. It discriminates speaker embeddings in a batch. Not only the $i$-th student embedding is pulled towards the corresponding teacher embedding, but also other student embeddings are pushed away from it. The underlying assumption is that any two samples in the batch are unlikely to be from the same speaker. This is reasonable considering the relatively small batch size with respect to the large speaker population.

The contrastive loss was proposed in SSL\cite{he2020momentum,ding2020learning}, where both $v_t^i$ and $v_s^i$ come from the same extractor.
The contrastive loss is also closely related to angular softmax loss\cite{deng2019arcface,liu2019large}. The major difference is that when using the angular softmax loss, $v_t^i$ is part of the model parameters. It is the last linear layer added to the embedding extractor for supervised training. This linear layer can be regarded as a speaker lookup table of fixed size, which is randomly initialized.

\section{Experiments}
\subsection{Configurations}
\subsubsection{Data\label{sec:data}}
Experiments of label-free KD are conducted on several large-scale datasets for speaker recognition. As listed in Table \ref{tab:datasets}, VoxCeleb2\cite{nagrani2017voxceleb} and CNCeleb\cite{li2022cn} are the two largest public datasets for SR with the languages of English and Chinese, respectively. In addition to that, several datasets for other speech tasks are incorporated as unlabelled data for KD pretraining. They are WenetSpeech\cite{zhang2021wenetspeech} for Chinese speech-to-text, Gigaspeech\cite{chen2021gigaspeech} for English speech-to-text and Voxlingua107\cite{valk2021voxlingua107} for language recognition. They are much larger than the aforementioned SR datasets, except that speaker labels are missing.

The training data are preprocessed to fit SR tasks. We keep only the speech samples longer than 2 seconds. 80-dim filterbanks are then extracted, followed by voice activity detection and mean normalization with a sliding window up to 3 seconds.

\begin{table}[htbp]
    \centering
    \caption{Summary of training data after pre-processing}
    \begin{tabular}{c||c|c|c|c}
    \toprule
        & \#utt & \#spk & duration & language \\
    \hline
    \hline
       VoxCeleb2 & 1.09M & 5995 & 2234h & most English \\
       CNCeleb1+2 & 560K & 2792 & 1214h & Chinese  \\
       WenetSpeech & 13M & - & 6602h & Chinese \\
       GigaSpeech  & 7.6M & - & 7959h & English \\
       Voxlingua107 & 2.5M & - & 6173h & multiple \\
    \hline
        total & 24.75M & - & 24182h & multiple \\
    \bottomrule
    \end{tabular}
    \label{tab:datasets}
\end{table}

\subsubsection{Model}
Computational cost is a major concern in practice. Table \ref{tab:model} compares several state-of-the-art neural networks in terms of model size and real time factor (RTF). The RTF is calculated on an Intel(R) Xeon(R) CPU E5-2630 v4 @ 2.20GHz with input samples of 10sec. As shown in Table \ref{tab:model}, state-of-the-art speaker recognition models like SE-ResNet34\cite{chung2020defence} and ECAPA-TDNN\cite{desplanques2020ecapa} are computationally intensive. The most recent Transformer-based models\footnote{\text{https://huggingface.co/microsoft/wavlm-base-sv}}\footnote{\text{https://huggingface.co/microsoft/wavlm-large}} trained by SSL\cite{baevski2020wav2vec,chen2021wavlm,chen2021large} are much heavier than the aforementioned. In comparison, the classical x-vector extractor\cite{snyder2018x} is applicable due to its relatively small model size and low RTF. The light-weight model\footnote{\label{note:mobilenet}\text{https://github.com/pytorch/vision/blob/main/torchvision/models}}, MobileNet3\cite{howard2019searching} is employed from image classification to speaker recognition. It is more light-weight than the x-vector extractor.

In our experiments, SE-ResNet34 is adopted as the teacher model. It consists of 34 stacked squeeze-excitation blocks\footnote{\text{https://github.com/clovaai/voxceleb\_trainer/tree/master/models}}. The MobileNet3 and the x-vector extractor are employed as the student models.
Modifications to MobileNet3 are limited: (1) reducing the number of input channels from 3 to 1; (2) changing the last classification dimension to the embedding dimension (256-dim). Note that we use the channel-wise global average pooling for MobileNet3.

\begin{table}[htbp]
    \centering
    \caption{Computational comparison between neural networks}
    \begin{tabular}{c||c|c||c|c}
    \toprule
        & \#para. & RTF  & teacher & student \\
    \hline
    \hline
       MobileNet3 (big) & 1.9M & 0.002  &  & \checkmark \\
       x-vector & 3.6M & 0.003 & & \checkmark \\
    %   \textbf{thin-resnet34}{\color{red} TODO} &  & & & \checkmark \\
    %   fast-resnet34 & & & & \checkmark \\
       SE-ResNet34 & 24.2M & 0.049 & \checkmark &  \\
    \hline
       ECAPA-TDNN & 15.6M & 0.014 & \\
       Transformer(base) & 94.3M & 0.083 & & \\
       Transformer(large) & 315M & 0.276 & & \\
    \bottomrule
    \end{tabular}
    \label{tab:model}
\end{table}
% \begin{table}[htbp]
%     \centering
%     \caption{Computational comparison between speaker models}
%     \begin{tabular}{c||c|c|c|c}
%     \toprule
%         & Size & RTF & FLOPs & Choice \\
%     \hline
%     \hline
%       mobilenet3 & 1.9M & 0.0029 & 0.84G & student \\
%       shufflenet2 & 3.6M & 0.0026 & 5.6G & student \\
%       xvector & 5.8M & 0.005 & 2G & student \\
%       se-resnet34 & 24M & 0.055 & 140G & teacher \\
%       ecapa & 15.6M & 0.013 & 28G & - \\
%       wavlm &  & & \\
%     \bottomrule
%     \end{tabular}
%     \label{tab:backend_cnceleb}
% \end{table}

\subsubsection{Model Training and Evaluation}
% In the supervised training with speaker labels, we use AAM-softmax loss with a margin of 0.3 and a scale of 30. The SGD optimizer is adopted with the learning rate exponentially decayed from 0.1 to 0.01. We use a batch of 1024 speech segments with duration of 2-3s. Only spectrum augmentation is applied to the input data in all experiments. We keep the settings the same for both knowledge distillation and label-based training.

For simplicity, the KD training, supervised training and fine-tuning share almost the same configurations in all of our experiments. The stochastic gradient descent (SGD) optimizer is adopted with the learning rate exponentially decayed from 0.1 to 0.01. The models are trained using two NVIDIA V100 GPUs with 32GB memory. Each GPU is allocated with a batch of 512 speech segments (2s-3s). The training goes through the whole training data repeatedly for 15 times. 
A subset of 200-hour speech samples are randomly selected for every epoch. Only spectrum augmentation\cite{park2019specaugment} is applied.
In the supervised training with speaker labels, AAM-softmax\cite{liu2019large} loss is adopted with the scalar set to 30. Similar to the curriculum learning proposed in \cite{chung2020defence}, the angular margin of AAM-softmax loss is set to 0 for the first 30 epochs and then increases to $0.3$.

After training, 256-dim embeddings extracted from the neural networks are length-normalized\cite{garcia2011analysis}, followed by cosine scoring to produce verification scores.
Equal error rate (EER) is used for performance evaluation. It balances between false alarm errors and false reject errors.

\subsection{In-domain Distillation}

We first conduct in-domain distillation on VoxCeleb to verify the effectiveness of contrastive loss. Both teacher and student models are trained with the training data of VoxCeleb2 and evaluated on the test data of VoxCeleb1. Table 3 shows the results. \textit{AAM} denotes the supervised training by AAM softmax loss that requires speaker labels. The others denote label-free distillation as described in Section \ref{sec:distillation_loss}. Clearly the contrastive loss outperforms conventional distillation loss like \textit{MSE} and \textit{COS}. It can be even better than \textit{AAM} training on MobileNet3. Its superiority is partially due to the well-trained teacher model, SE-ResNet34, which achieves the EER of 1.0\%. These results show that the contrastive loss can effectively distill knowledge from the teacher model to improve the students.
\begin{table}[htbp]
    \centering
    \caption{Comparison between supervised training with AAM-softmax loss and label-free knowledge distillation}
    \begin{tabular}{c||c||c|c|c}
    \toprule
       model & AAM  & MSE & COS & contrastive \\
    \hline
    \hline
       MobileNet3 & 4.06 & 7.82 & 4.96 & \textbf{3.77}  \\
       % ShuffleNet2 & 6.19 & 6.77 & 5.24 & \textbf{4.11}  \\
       x-vector & \textbf{2.71} & 4.52 & 4.08 & 2.86 \\
    %   thin-resnet34 & \textbf{2.81} & 
    \hline
       SE-ResNet34 & 1.00 & - & - & - \\
    \bottomrule
    \end{tabular}
    \label{tab:indomain_distill}
\end{table}

\subsection{Out-of-domain Distillation}
In the previous experiment, training of the teacher model and distillation for the student models are performed on the same data. To investigate the robustness of the proposed distillation method on out-of-domain data, the CNCeleb dataset is used. The teacher remains the same (trained on VoxCeleb2), while the students are trained on both CNCeleb1+2 and VoxCeleb2. Table \ref{tab:transfer_cnceleb} shows the evaluation results on CN-Celeb. The effectiveness of contrastive loss remains very significant in this case of out-of-domain distillation.

For MobileNet3, contrastive loss outperforms \textit{MSE} and \textit{COS} loss by absolute margins of $7.66\%$ and $4.77\%$, respectively. It is noticeably better than supervised training with AAM-sofmax loss (\textit{AAM}). For x-vector, contrastive loss is comparable to \textit{AAM}(12.98\% versus 11.71\%). The superiority of contrastive loss suggests that label-free KD can also be used for unsupervised domain adaptation of the front-end, e.g., unsupervisedly pretraining on a large amount of target domain data.
\begin{table}[htbp]
    \centering
    \caption{transfer to CNCeleb without finetuning}
    \begin{tabular}{c||c||c|c|c}
    \toprule
       model & AAM & MSE & COS & contrastive \\
    \hline
    \hline
       MobileNet3 & 16.74 & 21.03 & 18.14 & \textbf{13.37}  \\
       % ShuffleNet2 & 17.29 & 20.10 & 19.35 & \textbf{13.50}  \\
       x-vector & \textbf{11.71} & 18.59 & 18.05 & 12.98 \\
    \bottomrule
    \end{tabular}
    \label{tab:transfer_cnceleb}
\end{table}

\subsection{Large-scale Distillation and Fine-tuning}

The main goal of this study is about effective pre-training of light-weight models on large-scale unlabelled data collected from diverse sources, as described in Section \ref{sec:data}. Pretraining of student models are computationally affordable, due to their light-weight property. With two V100 GPUs, it costs less than two days to pretrain a student model on the combined data of 24,182 hours.

Table \ref{tab:large_distill_voxceleb} evaluates the student models after large-scale distillation and finetuning. The evaluation is conducted on the VoxCeleb1 test data.  The proposed contrastive loss still works well on the x-vector model.
The finetuning on x-vector gives significant improvements, e.g., EER is reduced from 2.95\% to 2.09\% for x-vector with contrastive loss. It indicates the superior network design of the x-vector extractor.
Another observation is that the performance gap between different loss shrinks as the unlabelled data become very large.

For MobileNet3, the unsupervised KD is still very useful, e.g., reducing EER from 4.06\% (no distillation, trained by AAM loss) to 2.90\% (distillation applied). However, the finetuning with AAM loss on VoxCeleb2 does not introduce further improvements, e.g., it is degraded from 2.90\% to 3.15\% in terms of EER. The performance result is an interpolation between the unsupervised KD pretraining ($2.90\%$) and the supervised training by AAM loss ($4.06\%$). Due to the relatively poor performance of MobileNet3 trained by AAM loss, finetuning by AAM loss is not beneficial to the system performance. In this regard, it is suggested to skip the finetuning of MobileNet3.

We also compare the proposed KD with other related studies that used SSL for x-vector and ResNet. These studies attempt to train a speaker model from scratch without any external supervision. To improve the performance, some use extensive data augmentation methods and tricky training strategies, which are quite complicated in practice and difficult to scale up to large-scale data. Moreover, the performance of SSL on light-weight models is far from satisfactory, compared to our simple knowledge distillation method. It would be unnecessary to train from scratch without considering many excellent teacher models that are publicly available.
% \begin{table}[htbp]
%   \centering
%   \caption{Large-scale distillation and finetuning. Models are evaluated on VoxCeleb1}
%   \begin{tabular}{c||c||c|c|c}
%     \toprule
%      & AAM & MSE & COS & contrastive \\
%     \hline
%     \hline
%      \multicolumn{5}{l}{\textbf{distillation}} \\  
%     \midrule
%      MobileNet3 & 4.06 & 3.34 & 3.20 & \textbf{2.9} \\
%      % ShuffletNet2 & 6.19 & 3.56 & 3.19  & \textbf{2.89} \\
%      x-vector & \textbf{2.71} & 3.19 & 3.01 & 2.95 \\
%      \midrule
%     \multicolumn{5}{l}{\textbf{+ finetuning on VoxCeleb2}} \\  
%      \midrule
%      MobileNet3 & - & 4.52 & 3.72 & \textbf{3.15} \\
%      % ShuffletNet2 & - & 5.79 & 5.15 & \textbf{4.93} \\
%      x-vector & - & 2.46 & 2.21 & \textbf{2.09} \\
%      \midrule
%     \multicolumn{5}{l}{\textbf{Other related work}} \\
%      \midrule
%     \multicolumn{4}{l}{moco+ProtoNCE+x-vector\cite{xia2021self}} & 8.23 \\
%     \multicolumn{4}{l}{siamese+aug+ResNet\cite{sang2021self}} & 6.99 \\
%     \multicolumn{4}{l}{bootstrap + ResNet\cite{mun2021bootstrap}} & 6.42 \\
%     \multicolumn{4}{l}{moco+x-vector\cite{ding2020learning} (tune on VoxCeleb1+2)} & 2.40 \\
%     \bottomrule
%   \end{tabular}
%   \label{tab:large_distill_voxceleb}
% \end{table}
\begin{table}[htbp]
  \centering
  \caption{Large-scale distillation and finetuning. Models are evaluated on VoxCeleb1}
  \begin{tabular}{c||c|c|c}
    \toprule
      & MSE & COS & contrastive \\
    \hline
    \hline
     \multicolumn{4}{l}{\textbf{distillation}} \\  
    \midrule
     MobileNet3 & 3.34 & 3.20 & \textbf{2.90} \\
     % ShuffletNet2 & 6.19 & 3.56 & 3.19  & \textbf{2.89} \\
     x-vector & 3.19 & 3.01 & 2.95 \\
     \midrule
    \multicolumn{4}{l}{\textbf{+ finetuning on VoxCeleb2}} \\  
     \midrule
     MobileNet3 & 4.52 & 3.72 & \textbf{3.15} \\
     % ShuffletNet2 & - & 5.79 & 5.15 & \textbf{4.93} \\
     x-vector & 2.46 & 2.21 & \textbf{2.09} \\
     \midrule
    \multicolumn{4}{l}{\textbf{supervised training without distillation}} \\
     \midrule
     \multicolumn{3}{l}{MobileNet3} & 4.06 \\
     \multicolumn{3}{l}{x-vector} & 2.71 \\
     \midrule
    \multicolumn{4}{l}{\textbf{Other related work}} \\
     \midrule
    \multicolumn{3}{l}{moco+ProtoNCE+x-vector\cite{xia2021self}} & 8.23 \\
    % \multicolumn{3}{l}{siamese+aug+ResNet\cite{sang2021self}} & 6.99 \\
    \multicolumn{3}{l}{bootstrap + ResNet\cite{mun2021bootstrap}} & 6.42 \\
    \multicolumn{3}{l}{moco+x-vector\cite{ding2020learning}} & 2.40 \\
    \bottomrule
  \end{tabular}
  \label{tab:large_distill_voxceleb}
\end{table}

\begin{table}[htbp]
  \centering
  \caption{Large-scale distillation and finetuning. Models are evaluated on CNCeleb1}
  \begin{tabular}{c||c|c|c}
    \toprule
      & MSE & COS & contrastive \\
     \hline
     \hline
      \multicolumn{4}{l}{\textbf{distillation}} \\  
     \midrule
      MobileNet3 & 15.72 & 16.84 & \textbf{12.74} \\ %
      % ShuffleNet2 & 17.29 & 16.74 & 17.37 & \textbf{13.09} \\
      x-vector & 15.97 & 15.97 & 12.22  \\ % 
     \midrule
    \multicolumn{4}{l}{\textbf{+ finetuning on CNCeleb1+2}} \\
     \midrule
      MobileNet3 & 16.33 & 14.90 & \textbf{12.82} \\
      % ShuffleNet2 & - & 18.16 & 18.10 & \textbf{16.51} \\
      x-vector & 10.46 & 10.12 & \textbf{9.43} \\
     \midrule
    \multicolumn{4}{l}{\textbf{supervised training without distillation}} \\
     \midrule
     \multicolumn{3}{l}{MobileNet3} & 16.74 \\
     \multicolumn{3}{l}{x-vector} & 11.71 \\
    \midrule
    \multicolumn{4}{l}{\textbf{Other related work}} \\
    \midrule
    \multicolumn{3}{l}{x-vector + dnf\cite{cai2020deep}} & 14.22 \\
    \multicolumn{3}{l}{x-vector + aug\cite{tian2022royalflush}} & 13.74 \\
    \multicolumn{3}{l}{ResNet + aug\cite{tian2022royalflush}} & 9.32 \\
    \multicolumn{3}{l}{ECAPA-TDNN + aug\cite{tian2022royalflush}} & \textbf{8.82} \\
    
    \bottomrule
  \end{tabular}
  \label{tab:large_distill_cnceleb}
\end{table}

Table \ref{tab:large_distill_cnceleb} summaries the evaluation results on CNCeleb, where the teacher model is trained on VoxCeleb2. The fine-tuned x-vector extractor can still significantly benefit from large-scale distillation based on the contrastive loss, e.g., EER reduced from $11.71\%$(no distillation) to $9.43\%$(distillation+finetuning). The finetuned x-vector outperforms \textit{x-vector+aug} and is even competitive to \textit{ResNet+aug}. Similar to experiments on VoxCeleb, MobileNet3 after distillation does no benefit from the subsequent finetuning by AAM loss.

\subsection{Comparison among Light-weight Models}
The choice of student models in KD should also be reckoned with. As shown in Table \ref{tab:indomain_distill}-\ref{tab:large_distill_cnceleb}, the  performance varies between the student models. The classical x-vector extractor in general outperforms the MobileNet3, especially after large-scale distillation and finetuning. A possible reason is that the MobileNet3 is originally designed for image classification. The network structures need more modifications to fit for speech tasks. It also indicates that the x-vector extractor is quite well-designed for SR. We hope in the future there could be more research on designing light-weight speaker extractors so as to further unleash the power of KD.

% \vspace{-1mm}
\section{Conclusion}\label{conclusions}
Light-weight speaker extractors are highly desired in practical applications. This research aims to improve their performance by efficiently incorporating large amount of unlabelled data for training. It is achieved by leveraging the contrastive loss from self-supervised learning for label-free knowledge distillation. 
Several light-weight neural networks are employed as speaker extractors to evaluate the effectiveness of label-free knowledge distillation.
Experiments with VoxCeleb and CNCeleb demonstrate the superiority of contrastive loss over both the conventional distillation loss and the state-of-the-art angular-based classification loss.
Unlabelled data are scaled up by combining several large-scale speech datasets from diverse sources. The large-scale knowledge distillation is effective in boosting the performance of light-weight speaker extractors.

\bibliographystyle{IEEEtran}

\bibliography{template}

% \begin{thebibliography}{9}
% \bibitem[1]{Davis80-COP}
%   S.\ B.\ Davis and P.\ Mermelstein,
%   ``Comparison of parametric representation for monosyllabic word recognition in continuously spoken sentences,''
%   \textit{IEEE Transactions on Acoustics, Speech and Signal Processing}, vol.~28, no.~4, pp.~357--366, 1980.
% \bibitem[2]{Rabiner89-ATO}
%   L.\ R.\ Rabiner,
%   ``A tutorial on hidden Markov models and selected applications in speech recognition,''
%   \textit{Proceedings of the IEEE}, vol.~77, no.~2, pp.~257-286, 1989.
% \bibitem[3]{Hastie09-TEO}
%   T.\ Hastie, R.\ Tibshirani, and J.\ Friedman,
%   \textit{The Elements of Statistical Learning -- Data Mining, Inference, and Prediction}.
%   New York: Springer, 2009.
% \bibitem[4]{YourName17-XXX}
%   F.\ Lastname1, F.\ Lastname2, and F.\ Lastname3,
%   ``Title of your INTERSPEECH 2022 publication,''
%   in \textit{Interspeech 2022 -- 23\textsuperscript{rd} Annual Conference of the International Speech Communication Association, September 18-22, Incheon, Korea, Proceedings, Proceedings}, 2022, pp.~100--104.
% \end{thebibliography}

\end{document}